\documentclass[conference]{IEEEtran}
\IEEEoverridecommandlockouts
\usepackage{graphicx}
\usepackage{cite}
\usepackage{amsmath}
\usepackage{amssymb}
\usepackage{verbatim}
\usepackage{multirow}
\usepackage{diagbox}
\usepackage{booktabs}

\usepackage{algpseudocode}
\usepackage[ruled]{algorithm2e}
\usepackage{algorithm2e}

\usepackage[top=0.71in, bottom=1in, left=0.625in, right=0.625in]{geometry}

\def\ba{{\boldsymbol a}}
\def\bs{{\boldsymbol s}}

\def\by{{\boldsymbol y}}

\def\bw{{\boldsymbol w}}
\def\bR{{\boldsymbol R}}

\def\bX{{\boldsymbol X}}
\def\bW{{\boldsymbol W}}
\def\bI{{\boldsymbol I}}
\def\bH{{\boldsymbol H}}
\def\bF{{\boldsymbol F}}

\def\BibTeX{{\rm B\kern-.08em{\sc i\kern-.025em b}\kern-.08em
    T\kern-.1667em\lower.7ex\hbox{E}\kern-.125emX}}
\begin{document}

\title{Sustainable LSTM-Based Precoding for RIS-Aided mmWave MIMO Systems with Implicit CSI
\thanks{This work was supported in part by the Academia Sinica (AS) under Grant 235g Postdoctoral Scholar Program, and in part by the National Science and Technology Council (NSTC) of Taiwan under Grant 113-2926-I-001-502-G, 113-2221-E-110-059-MY3, and 114-2218E-110-005.}
}
\author{\IEEEauthorblockN{Po-Heng Chou$^{1}$, Jiun-Jia Wu$^{2}$, Wan-Jen Huang$^{2}$, and Ronald Y. Chang$^{1}$}
\IEEEauthorblockA{$^{1}$Research Center for Information Technology Innovation (CITI), Academia Sinica (AS), Taipei 11529, Taiwan\\
$^{2}$Institute of Communication Engineering (ICE), National Sun Yat-sen University (NSYSU), Kaohsiung 80424, Taiwan\\
E-mails: d00942015@ntu.edu.tw, wl021156484@gmail.com, wjhuang@faculty.nsysu.edu.tw, rchang@citi.sinica.edu.tw}
\vspace{-0.45in}
}
\maketitle

\begin{abstract}
In this paper, we propose a sustainable long short-term memory (LSTM)-based precoding framework for reconfigurable intelligent surface (RIS)-assisted millimeter-wave (mmWave) MIMO systems. Instead of explicit channel state information (CSI) estimation, the framework exploits uplink pilot sequences to implicitly learn channel characteristics, reducing both pilot overhead and inference complexity. Practical hardware constraints are addressed by incorporating the phase-dependent amplitude model of RIS elements, while a multi-label training strategy improves robustness when multiple near-optimal codewords yield comparable performance. Simulations show that the proposed design achieves over 90\% of the spectral efficiency of exhaustive search (ES) with only 2.2\% of its computation time, cutting energy consumption by nearly two orders of magnitude. The method also demonstrates resilience under distribution mismatch and scalability to larger RIS arrays, making it a practical and energy-efficient solution for sustainable 6G wireless networks.
\end{abstract}

\IEEEpeerreviewmaketitle
%\vspace{-0.1in}

\section{Introduction}
Reconfigurable intelligent surface (RIS) technology~\cite{You2025, Lu2025, Abeywickrama2020TWC, Abeywickrama2020ICC, Ruijin2023, Chou2024IWCL, Chou2024Globecom, Huang2024PIMRC, Masoud2025, Liu2023, Jiang2021, Sheen2021, Ge2021, Ozdogan2020} has recently emerged as a cost-effective solution for improving both spectral and energy efficiency in millimeter wave (mmWave) multiple-input multiple-output (MIMO) systems~\cite{You2025}. By acting as a passive alternative to relay systems~\cite{Chou2015}, RIS mitigates blockage and extends mmWave coverage without active RF chains. Each reflective element adjusts the amplitude and phase of incident signals through a software-defined controller, enabling energy-efficient beam shaping. The physical implementation of RIS has been well studied~\cite{Lu2025, Abeywickrama2020TWC, Abeywickrama2020ICC}, but optimizing both active beamforming and passive reflection remains highly challenging due to non-convexity~\cite{Ruijin2023, Huang2024PIMRC, Masoud2025}. Traditional exhaustive search (ES) approaches are computationally prohibitive for real-time deployment, especially under the practical phase-dependent amplitude model~\cite{Abeywickrama2020TWC, Abeywickrama2020ICC}, which most prior works neglect~\cite{Liu2023, Jiang2021, Sheen2021, Ge2021, Ozdogan2020}.

To overcome these challenges, deep learning (DL)-based approaches have been explored for RIS optimization. Surveys such as~\cite{Ozpoyraz2022} show that DL methods reduce pilot and training overhead compared with conventional designs. For example, deep reinforcement learning (DRL) has been adopted for adaptive RIS phase configuration~\cite{Wen2020}, while long short-term memory (LSTM) models~\cite{Liu2021,LSTM2020} exploit temporal CSI correlations for improved beam prediction. Thanks to their memory cells and gating structures, LSTMs provide stronger capability than convolutional neural networks (CNNs) in capturing long-range temporal dependencies of pilot signals, which are critical for RIS precoding accuracy.

More recently, the sustainability of AI-aided wireless systems has become a key research focus. RIS offers passive, ultra-low power operation that improves both spectral and energy efficiency by eliminating active RF chains~\cite{Yang2022TWC}. In parallel, the concept of Green Artificial Intelligence (AI) emphasizes lightweight, inference-efficient models for 6G edge deployments~\cite{Zhang2024MCOM}. The proposed LSTM-based CSI-free precoding framework aligns with this vision, sustaining over $90\%$ of ES spectral efficiency while significantly lowering pilot overhead and inference complexity, thereby minimizing the energy footprint. In addition, our framework requires one-time offline training, with inference dominating deployment cost, which makes it practical for energy-efficient large-scale systems.

However, explicit channel state information (CSI) estimation at RIS remains challenging due to its passive nature. Several works~\cite{Liu2023, Jiang2021, Sheen2021, Ge2021, Ozdogan2020} have adopted learning-based approaches that bypass CSI estimation by mapping received pilot signals or location features directly to RIS configurations. In our framework, the uplink pilot sequence is obtained under a fixed reference RIS state (such as random or default), ensuring the input feature does not implicitly encode the target RIS configuration predicted. This design avoids the circularity concern discussed in~\cite{Jiang2021} and allows the LSTM to learn the mapping from unbiased pilot information to near-optimal RIS settings.

To further improve robustness, we adopt a multi-label training strategy~\cite{Zhang2014}. Instead of marking only the single best ES codeword as the positive label, we include multiple near-optimal codewords that achieve spectral efficiencies within $0.5$ dB of the ES optimum. This design reflects the non-convexity of RIS optimization and prevents the model from overfitting to a single ES outcome. It also enhances generalization against distribution mismatch, addressing concerns raised in~\cite{Ozpoyraz2022, Liu2021}. 

In addition to these technical aspects, practical deployment scenarios such as vehicular networks, UAV communications, and dense urban environments further highlight the necessity of RIS-aided solutions for 6G and beyond. While alternating optimization (AO) has been widely adopted, its high computational complexity limits scalability under fast-varying channels. DL-based methods not only reduce pilot overhead but also provide robustness against mobility and non-stationary environments, making them suitable for real-time use~\cite{Ozpoyraz2022}. Moreover, the passive nature of RIS is consistent with the vision of Green AI~\cite{Zhang2024MCOM}, where lightweight and inference-efficient models are critical for sustainable edge deployment.

Our contributions are highlighted as follows:
\begin{itemize}
    \item To the best of our knowledge, this is the first work to jointly account for (i) the practical phase–amplitude coupling constraint of RIS hardware~\cite{Abeywickrama2020TWC, Abeywickrama2020ICC} and (ii) CSI-free operation, within a unified LSTM-based precoding framework tailored for mmWave MIMO systems.
    \item We construct a Kronecker-structured DFT codebook that captures both azimuth and elevation features. Leveraging LSTM’s gating mechanism for temporal pilot correlation, the proposed model consistently outperforms CNN baselines in both spectral efficiency and inference latency, while supporting scalability to larger RIS arrays.
    \item We introduce a multi-label classification approach~\cite{Zhang2014}, where multiple near-optimal codewords (within $0.5$ dB of the ES optimum) are treated as positive labels. This design mitigates overfitting to a single ES label and enhances robustness against distribution mismatch, addressing concerns highlighted in~\cite{Ozpoyraz2022, Liu2021}.
    \item The proposed LSTM framework achieves over $90\%$ of the spectral efficiency of ES with only $2.2\%$ of its inference complexity, translating into nearly two orders of magnitude energy savings. This aligns with the vision of green AI and sustainable 6G deployments~\cite{Zhang2024MCOM}.
\end{itemize}

\section{System and Channel Models}

\begin{figure}[tb]
\centering
{\includegraphics[width=0.38\textwidth]{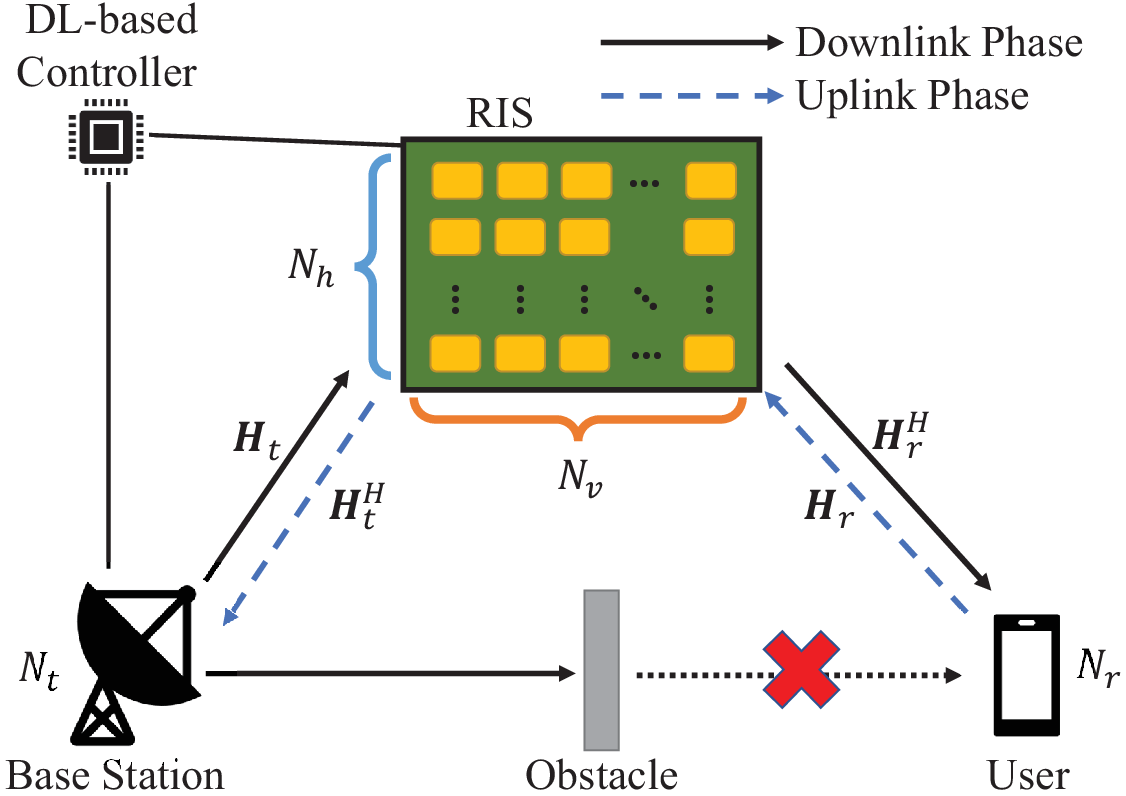}
\caption{The RIS-assisted mmWave MIMO system.}\label{System_Model}}
\vspace{-0.2in}
\end{figure}

Consider a RIS-assisted MIMO system operated in the mmWave spectrum, where a BS with $N_t$ antennas communicates with a user equipped with $N_r$ antennas. The direct path is assumed blocked; therefore, the transmission is assisted by an RIS comprising $N = N_h \times N_v$ passive reflecting elements arranged in a uniform planar array (UPA), with $N_h$ and $N_v$ denoting the number of elements along the horizontal and vertical dimensions, respectively, as illustrated in Fig.~\ref{System_Model}.

During the data downlink phase, the signal transmitted from the BS arrives at the user after being reflected by the RIS. Let $\bs \in \mathbb{C}^{N_s \times 1}$ represent the downlink data streams with covariance matrix $E[\bs \bs^H]=\frac{P}{N_{s}}\bI_{N_{s}}$, where $P$ is the total transmission power. The equivalent baseband signal at the user is given by
\begin{equation}\label{eq:rx_sig}
\by = \bH_{r}^{H}\boldsymbol{\Psi}\bH_{t}\bF\bs + \bw,
\end{equation}
where $\boldsymbol{H}_{r}^{H} \in \mathbb{C}^{N_{r} \times N}$ and $\boldsymbol{H}_{t} \in \mathbb{C}^{N \times N_{t}}$ are the channel matrices from the RIS to the user and from the BS to the RIS, respectively, $\boldsymbol{\Psi} \in \mathbb{C}^{N \times N}$ is a diagonal matrix of the RIS response and $\boldsymbol{F} \in \mathbb{C}^{N_{t} \times N_{s}}$ is a precoding matrix at BS and $\bw\thicksim {\cal CN}(0, \sigma^{2}\bI_{N_{s}}) \in \mathbb{C}^{N_t \times 1}$ is the additive white Gaussian noise (AWGN) vector.
The effective channel is defined as $\boldsymbol{H}_{\text{eff}} = \boldsymbol{H}_r^H \boldsymbol{\Psi} \boldsymbol{H}_t$, representing the cascaded channel from the BS to the user via the RIS. To fully exploit the available spatial degrees of freedom, the number of data streams is set to $N_s = \text{rank}(\boldsymbol{H}_{\text{eff}})$.

Most existing work assumed that the CSI of the RIS-assisted MIMO system is perfectly known \cite{Chou2024IWCL, Chou2024Globecom}. However, it requires additional feedback of CSI from the receiver, or, under the assumption of channel reciprocity, performs channel estimations through the uplink.
In this work, it is assumed that the channel reciprocity holds in~\cite{Mishra2019} and~\cite{Jiang2021}.
By bypassing the explicit channel estimation stage, we aim to optimize the downlink transmission directly from the pilot sequences received from the uplink.
To avoid the circularity concern raised in~\cite{Jiang2021}, we clarify that the pilot sequence $\bR$ is obtained under a fixed reference RIS state (e.g., random or identity configuration), independent of the final $\boldsymbol{\Psi}$ to be predicted. This ensures that $\bR$ provides unbiased channel information without presupposing the optimal RIS configuration.

Let $\mathbf{X} = [\mathbf{x}(1), \ldots, \mathbf{x}(K)] \in \mathbb{C}^{N_r \times K}$ represent the uplink pilot matrix transmitted from the user, where each column $\mathbf{x}(k)$ is a 16-QAM modulated symbol. The pilot sequences are designed to be orthogonal across time, satisfying $E[\mathbf{x}(k)\mathbf{x}^H(k')] = \delta_{k,k'} \cdot \frac{P}{N_r} \mathbf{I}_{N_r}$. The received signal at the BS is given by
\begin{equation}
\bR = \left(\bH_r^H \boldsymbol{\Psi} \bH_t\right)^H \bX + \bW_t,
\end{equation}
where $\bW_t\in \mathbb{C}^{N_t \times K}$ is the AWGN of the reverse link.
Rather than estimating the effective channel matrix $\bH_{\text{eff}} = \bH_r^H \boldsymbol{\Psi} \bH_t$, the received pilot sequence $\bR$ is used directly as input to a DL-based model for the prediction of the RIS phase configuration.
During the inference phase, the predicted RIS configuration is applied in the downlink data transmission phase. %Due to channel reciprocity, the same RIS configuration is assumed to be valid for both uplink pilot training and downlink data transmission. 
The proposed approach avoids explicit CSI acquisition while retaining spatial and temporal characteristics relevant to RIS optimization.

Consider that both the BS and the user are equipped with ULAs, and channels are modeled under Rician fading with line-of-sight (LOS) dominance. The respective channel matrices are expressed as
\begin{align}
\bH_t \!\!&= \!\!\sqrt{L_t} \sum_{\ell=0}^{L} z_{t,\ell} [\ba^H_{N_h}(\phi_{h,\ell}) \otimes \ba^H_{N_v}(\phi_{v,\ell})] \ba_{N_t}(\theta_{t,\ell})
, \\
\bH_r^H \!\!&=\!\! \sqrt{L_r} \sum_{\ell=0}^{L} z_{r,\ell} \ba_{N_r}^H(\theta_{r,\ell}) [\ba_{N_v}(\varphi_{v,\ell}) \otimes \ba_{N_h}(\varphi_{h,\ell})],
\end{align}
%\left(\sqrt{\frac{K_t}{K_t + 1}} \overline{\bH}_t + \sqrt{\frac{1}{K_t + 1}} \widetilde{\bH}_t \right)
%\left(\sqrt{\frac{K_r}{K_r + 1}} \overline{\bH}_r^H + \sqrt{\frac{1}{K_r + 1}} \widetilde{\bH}_r^H \right)
where $L_t$ and $L_r$ represent path loss, $L$ is the number of non-line-of-sight (NLOS) paths, $z_{t,\ell}$ and $z_{r,\ell}$ are normalized path gains of two channels, $\otimes$ represents the Kronecker product, and the index $\ell=0$ indicates the LOS path. The path loss in dB is modeled as
\begin{align*}
L_\tau(\text{dB}) &= L_0(\text{dB}) - 10\xi_\tau \log_{1\tau}\left(\frac{d_\tau}{D_0}\right) + G_{\text{RIS}}(\text{dB}), 
\end{align*}
for $\tau\in\{t,r\}$, where $d_t$ and $d_r$ are distances from the BS and user to RIS, $\xi_t$ and $\xi_r$ are path loss exponents, $D_0$ is the reference distance, and $G_{\text{RIS}}$ is the RIS gain. For the LOS components, the normalized path gains are $z_{t,\ell}=\sqrt{\frac{K_t}{K_t + 1}}$ and $z_{r,\ell}=\sqrt{\frac{K_r}{K_r + 1}}$, with $K_t$ and $K_r$ being the Rician factors of two channels. On the other hand, the normalized path gains for NLOS paths are complex Gaussian distributed as
$z_{t,\ell} \sim \mathcal{CN}(0, \frac{1}{L(K_t + 1)})$ and $z_{r,\ell} \sim \mathcal{CN}(0, \frac{1}{L(K_r + 1)})$, respectively.
In the channel model, the beam-steering vector is defined as
$\ba_N(\theta) = \left[1, e^{-j\theta}, \ldots, e^{-j(N-1)\theta} \right]^T \in \mathbb{C}^{N \times 1}$, under the assumption of uniform linear arrays (ULAs) with half-wavelength antenna spacing $d=\lambda/2$.
%The angles $\theta$, $\phi$, and $\varphi$ represent AoD and AOA components in azimuth and elevation domains.
The parameters $\theta_{t,\ell}$ and $\theta_{r,\ell}$ represent the angle of departure (AOD) at the BS and the angle of arrival (AOA) at the user, respectively, for the $\ell$-th path. $\phi_{h,\ell}$ and $\phi_{v,\ell}$ represent the horizontal and vertical AOAs, and $\varphi_{h,\ell}$ and $\varphi_{v,\ell}$ represent the horizontal and vertical AODs of the $\ell$-th path at the RIS, respectively.

The RIS response matrix $\boldsymbol{\Psi}$ accounts for practical amplitude–phase dependency as
\begin{align}
\boldsymbol{\Psi} = \textrm{diag} (\boldsymbol{\Phi}) 
= \text{diag}(\beta_1 e^{j\psi_1}, \ldots, \beta_N e^{j\psi_N}),
\end{align}
where $\textrm{diag}(\cdot)$ denotes the vector-to-diagonal operator, 
$\boldsymbol{\Phi} \in \mathbb{C}^{N \times 1}$ is the RIS response vector, 
$\beta_n \in (0,1]$ is the amplitude and $\psi_n \in [-\pi, \pi]$ is the phase shift of the $n$-th element. 
In practice, $\beta_n$ varies with $\psi_n$, and we adopt the following model in~\cite{Abeywickrama2020TWC,Abeywickrama2020ICC}
\begin{align}
\beta_n = (1 - \beta_{\min}) \left( \frac{\sin(\psi_n - \psi_0) + 1}{2} \right)^\alpha + \beta_{\min}.
\end{align}
Explicitly incorporating this amplitude–phase coupling ensures the proposed design remains faithful to hardware constraints, 
which is critical for real-time and sustainable RIS-assisted 6G deployments.

\section{Problem Formulation}

The objective is to jointly design the transmit precoder and the RIS phase-shift configuration to maximize the spectral efficiency of the user for downlink transmission. 
The spectral efficiency is maximized by solving the following optimization problem
\begin{subequations}
\begin{align}
\max_{\boldsymbol{F}, \boldsymbol{\Psi}} \quad & \log_2 \left| \boldsymbol{I}_{N_s} + \frac{P}{\sigma^2 N_s} \boldsymbol{H}_{\text{eff}}^H \boldsymbol{F} \boldsymbol{F}^H \boldsymbol{H}_{\text{eff}} \right|, \\
\text{s.t.} \quad & \|\boldsymbol{F}\|_F^2 \leq N_s, \\
& [\boldsymbol{\Psi}]_{i,j} =
\begin{cases}
\beta(\psi_i) e^{-j \psi_i}, & i = j, \\
0, & i \neq j,
\end{cases}
\end{align}
\label{eq:optimization_full}
\end{subequations}
\hspace{-0.1in} where $|\cdot|$ represents the matrix determinant. The first constraint limits the transmission power, while the second enforces the diagonal structure of the RIS response with phase-dependent amplitude.
The problem is inherently non-convex due to the non-linear structure of $\boldsymbol{\Psi}$. A typical solution involves AO, which iteratively updates $\boldsymbol{F}$ and $\boldsymbol{\Psi}$ but suffers from high complexity and potential convergence to suboptimal points~\cite{Ruijin2023}. To overcome this, we adopt a DL-based approach that requires only a single forward pass. Conventional ES or AO-based solutions require complexity on the order of $\mathcal{O}(Q \cdot N_t N_r)$ or higher per channel realization, which grows almost linearly with both RIS size $N$ and codebook size $Q$. Such complexity is prohibitive for real-time deployment in practical 6G networks.

Given a fixed RIS configuration $\boldsymbol{\Psi}$, the optimal precoding matrix $\boldsymbol{F}_{\text{opt}}$ is obtained by performing singular value decomposition (SVD) on $\boldsymbol{H}_{\text{eff}} = \boldsymbol{U}\boldsymbol{\Lambda}\boldsymbol{V}^H$, where $\boldsymbol{U} \in \mathbb{C}^{N_r \times N_s}$ and $\boldsymbol{V} \in \mathbb{C}^{N_t \times N_s}$ are unitary matrices, and $\boldsymbol{\Lambda} = \text{diag}(\tau_1, \ldots, \tau_{N_s})$ contains the singular values.
The precoder is selected as $\boldsymbol{F}_{\text{opt}} = \boldsymbol{V}$, and the corresponding achievable rate simplifies to
\begin{equation}
R = \sum_{c=1}^{N_s} \log_2 \left(1 + \frac{P}{\sigma^2 N_s} \tau_c^2 \right).
\end{equation}
Thus, maximizing the rate reduces to optimizing the RIS response $\boldsymbol{\Psi}$ to maximize the singular values $\tau_c$ of the effective channel. 
The optimization remains non-convex primarily due to the nonlinear amplitude–phase coupling in $\boldsymbol{\Psi}$ and the discrete nature of codebook-based selection, which jointly make closed-form solutions analytically intractable.
This leads to the reformulated problem
\begin{subequations}
\begin{align}
\max_{\boldsymbol{\Psi}} \quad & \sum_{c=1}^{N_s} \log_2 \left(1 + \frac{P}{\sigma^2 N_s} \tau_c^2 \right), \\
\text{s.t.} \quad & [\boldsymbol{\Psi}]_{i,j} =
\begin{cases}
\beta(\psi_i) e^{-j \psi_i}, & i = j, \\
0, & i \neq j.
\end{cases}
\end{align}
\label{eq:optimization_reduced}
\end{subequations}

\hspace{-0.22in} However, the simplified problem~\eqref{eq:optimization_reduced} remains non-convex and analytically intractable.
To make the precoding process tractable, we select the optimal RIS configuration from a predefined codebook of candidate phase-shift vectors, where $Q$ is the output dimension (equal to the codebook size). Since the channel structure relies on beam-steering vectors, we generate the codebook using quantized azimuth and elevation angles.

The beam-steering vector for angle $\theta$ is defined as $\boldsymbol{g}(N, \theta) = [1, e^{-j\theta}, \ldots, e^{-j(N-1)\theta}]$. The azimuth and elevation angles are quantized as
\begin{align*}
\theta_{h,i} &= (i-1) Q_h, \quad Q_h = \frac{2\pi}{N_h}, \quad i = 1, \ldots, N_h, \\
\theta_{v,j} &= (j-1) Q_v, \quad Q_v = \frac{2\pi}{N_v}, \quad j = 1, \ldots, N_v.
\end{align*}
For the ideal case where $\beta_n = 1$, the codebook $\mathcal{P}_{\text{ideal}}$ is constructed as
\begin{equation}
\boldsymbol{p}_{i,j} = \boldsymbol{a}_{N_h}(\theta_{h,i}) \otimes \boldsymbol{a}_{N_v}(\theta_{v,j}), \quad \boldsymbol{p}_{i,j} \in \mathcal{P}_{\text{ideal}}.
\end{equation}

To incorporate hardware imperfections, we adjust each codeword in $\mathcal{P}_{\text{ideal}}$ by applying the amplitude model. Let $\boldsymbol{p} = [e^{j\theta_1}, \ldots, e^{j\theta_N}]^T$ be a codeword in the ideal case, and define the corresponding amplitude vector as $\Upsilon(\boldsymbol{p}) = [\beta_1, \ldots, \beta_N]^T$. The practical codebook becomes
\begin{equation}
\mathcal{P}_{\text{prac}} = \left\{ \Upsilon(\boldsymbol{p}) \odot \boldsymbol{p} \; : \; \boldsymbol{p} \in \mathcal{P}_{\text{ideal}} \right\},
\end{equation}
where $\odot$ represents the Hadamard (element-wise) product.

The number of candidates in $\mathcal{P}_{\text{prac}}$ equals $Q = N_h \times N_v$, consistent with the RIS array dimension. 
Hence, conventional AO or ES approaches require evaluating $\mathcal{O}(Q \cdot N_t N_r)$ operations per channel realization~\cite{Ruijin2023}, 
which grows nearly linearly with the RIS size $N$ and becomes prohibitive for real-time deployment. 
In contrast, the proposed LSTM outputs a near-optimal codeword in a single forward pass, 
reducing the computational cost by more than $97\%$ compared with ES. 
This drastic reduction not only enables real-time scalability but also translates into significant energy savings, 
a critical requirement for sustainable AI-driven RIS-assisted 6G systems.

\section{Long Short-Term Memory (LSTM) Model}
In this paper, we propose a pilot-based supervised LSTM model~\cite{LSTM2020} to predict the optimal phase configuration of RIS. The proposed LSTM directly learns from the temporal patterns of pilot signals, making it more robust to channel uncertainty.

\begin{figure}[t]
\centering
{\includegraphics[width=0.48\textwidth]{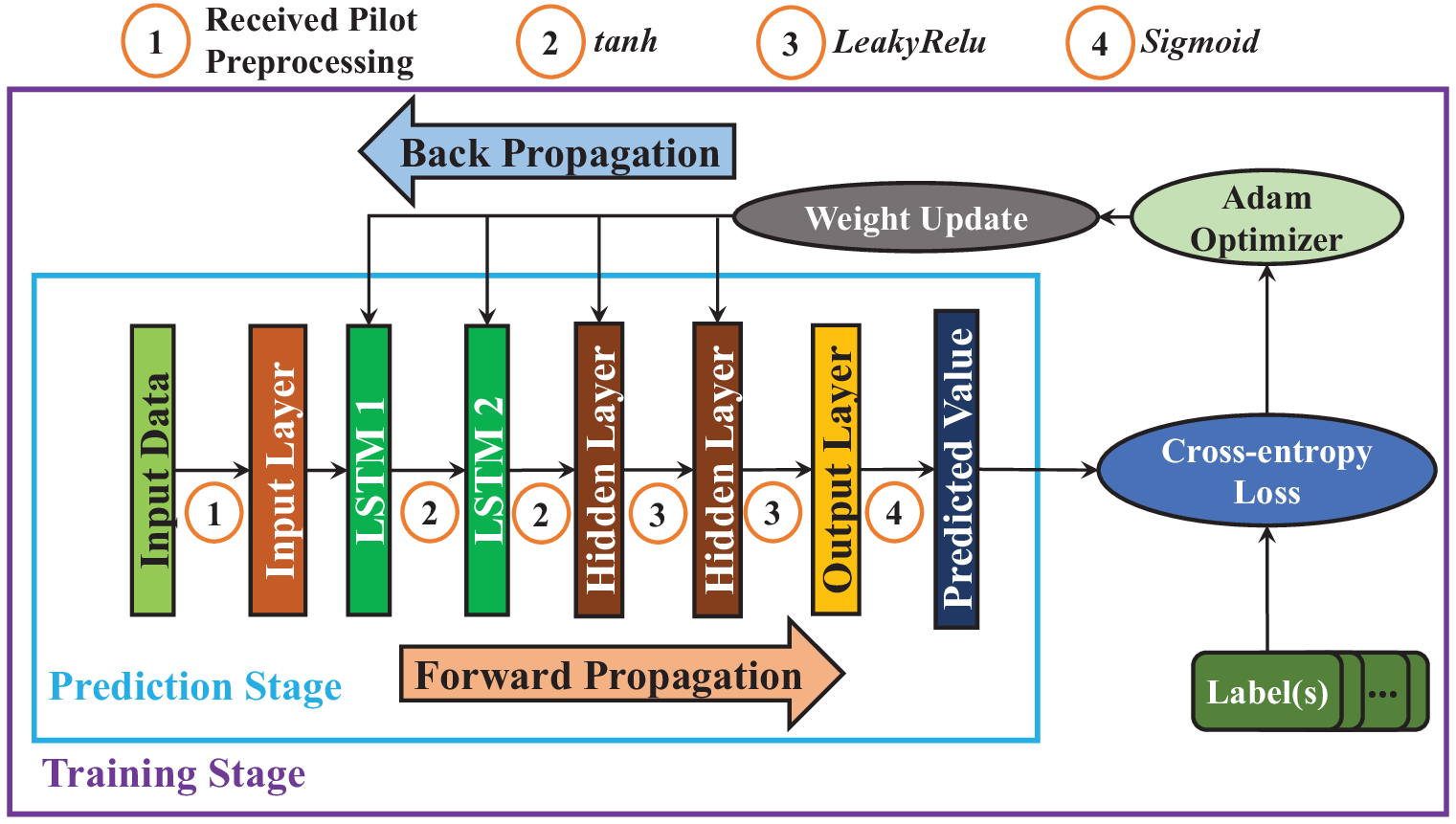}
\caption{LSTM model training flow chart.}\label{LSTM}}
\vspace{-0.2in}
\end{figure}

To address the problem of a large amount of CSI input data size caused by rapidly varying wireless channels, we utilize the received pilot signals $\bR$ at the BS as the input data for training the LSTM model, instead of using full CSI. Here, $k = 1, 2, \ldots, K$ represents the index of the pilot signal, and $c = 1, 2, \ldots, N_t$ represents the index of the BS antenna, while $N$ is the number of RIS elements. Before training, the input data is pre-processed as follows: (1) Perform Kronecker products on the pilot signal $\bR$ and the even-numbered columns of the RIS response vector $\boldsymbol{\Phi}$, which is given by $\bR \otimes \boldsymbol{\Phi}$. (2) Arrange $\bR \otimes \boldsymbol{\Phi}$ into a complex-valued vector. (3) Separate and rearrange the real and imaginary parts into a real-valued input vector. After preparing the training input, the supervised learning framework requires labels. The labels are determined via an ES to find the codeword in the codebook that maximizes spectral efficiency. The proposed LSTM model comprises an input layer, two LSTM layers, two hidden layers, and an output layer. The training phase involves two stages: forward propagation (prediction) and back-propagation (training), as illustrated in Fig.~\ref{LSTM}.

\textbf{Forward Propagation:} The input complex-valued pilot signal is first decomposed into real and imaginary parts, which are then concatenated into a real-valued vector. This input is fed into two sequential LSTM layers, each comprising 140 neurons with the \textit{tanh} activation function, to effectively capture temporal dependencies inherent in the pilot signal sequence. The LSTM outputs are subsequently passed through two fully connected hidden layers with 200 and 100 neurons, respectively. To ensure stable and efficient training, batch normalization is applied after each dense layer to normalize the mini-batch inputs to zero mean and unit variance, followed by the \textit{Leaky ReLU} activation function to introduce non-linearity and enhance feature representation. Finally, the network ends with an output layer consisting of $N$ neurons, corresponding to the codeword dimension of the DFT codebook~\cite{Yang2010}. A \textit{Sigmoid} activation function is applied to produce probability values in the range of [0, 1], and the predicted codeword vector is obtained by selecting elements whose output values exceed a predefined Sigmoid threshold, supporting multi-label classification~\cite{Zhang2014}.

\textbf{Back Propagation:} The model computes the loss between the predicted output and the true label using the cross-entropy loss function. One-hot encoding is applied to convert the label into a binary vector. The cross-entropy loss is defined as
\begin{align}
\mathcal{L} = -\frac{1}{B}\sum_{b=1}^{B}\sum_{q=1}^{Q}t_{b,q}\ln (y_{b,q}),
\end{align}
where $B$ is the batch size, $y_{b,q} \in [0,1]$ is the predicted value, and $t_{b,q}$ is the true label from one-hot encoding. To minimize the loss, the Adam optimizer is used due to its adaptive learning rate and robustness in handling complex data.

\begin{figure}[t]
\centering
{\includegraphics[width=0.49\textwidth]{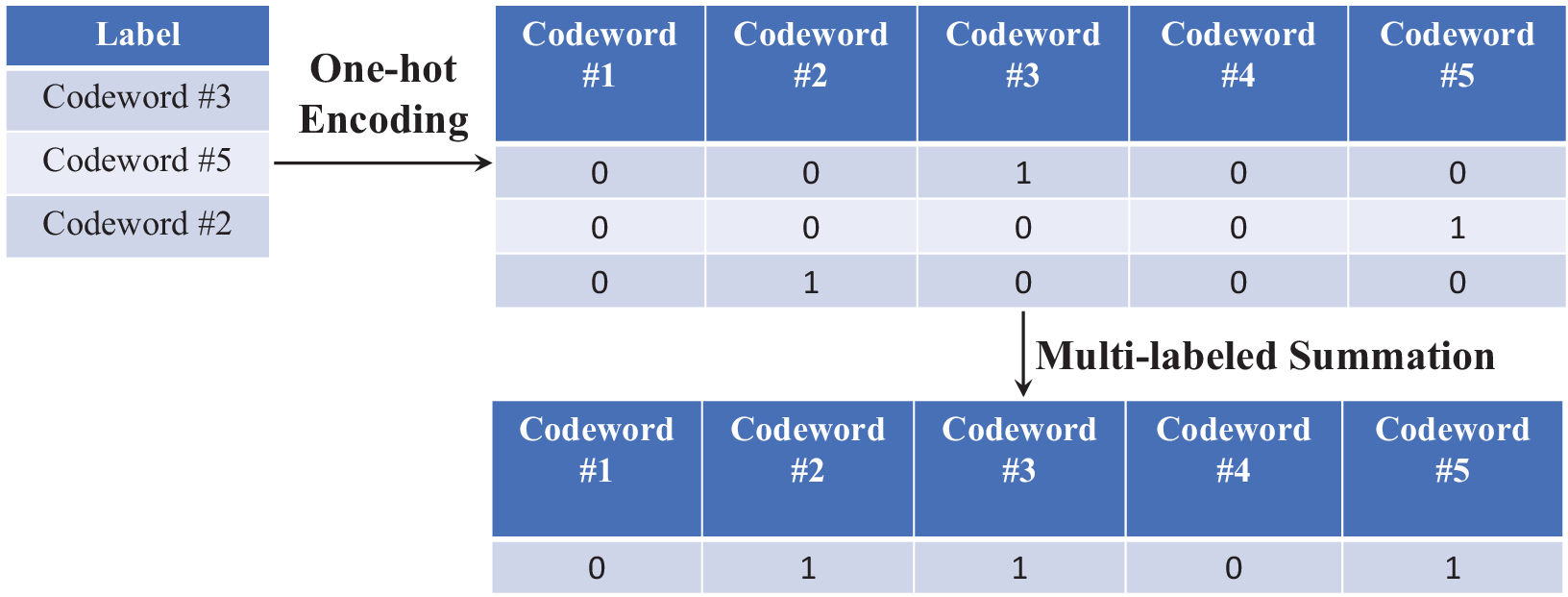}
\caption{Illustration of one-hot encoding for multi-labeled classification.}\label{onehot}}
\vspace{-0.2in}
\end{figure}

\textbf{Multi-Labeled Classification:} For single-labeled classification, the output layer of a supervised DL-based model typically employs the \textit{Softmax} activation function to convert raw outputs into a probability distribution over a predefined set of classes, selecting the class with the highest predicted probability. However, selecting the optimal codeword during training does not always guarantee the best performance at inference time. This discrepancy is primarily due to channel randomness and the non-convex nature of the precoding problem, where multiple codewords may yield similar or even better spectral efficiency depending on the deployment condition. To reflect this observation, also supported by our numerical results, we reformulate the RIS codeword selection task as a multilabel classification problem instead of a standard multiclass classification~\cite{Zhang2014}. The \textit{Sigmoid} function independently maps each output to a probability in the range [0, 1]. It allows the DL-based model to identify not only one codeword whose predicted probabilities exceed a preset threshold for RIS configuration.
For each output unit $z_i$ of the DL-based model, the \textit{Sigmoid} activation computes
\begin{equation}
\sigma(z_i) = \frac{1}{1 + e^{-z_i}}, \quad \text{for } i = 1, \ldots, Q.
\end{equation}
Since multiple near-optimal codewords may achieve similar performance, the multi-labeled approach enhances the flexibility of prediction.
The near-optimal codewords with predicted probabilities above the preset threshold are converted by the one-hot encoding and then summed.
For example, if the second, third, and fifth codewords have predicted probabilities above the preset threshold, these three codewords are near-optimal labels as shown in Fig.~\ref{onehot}.
It is worth noting that the proposed multi-label LSTM inference requires only a single forward pass with complexity $\mathcal{O}(K \cdot H)$, where $K$ is the pilot length and $H$ the hidden layer size. This is in stark contrast to the $\mathcal{O}(Q \cdot N_t N_r)$ operations required for ES and AO baselines, validating the scalability of the proposed framework for large RIS arrays. 

\section{Simulation Results}
In the simulation setup, we consider MIMO configurations where the BS is equipped with $N_t = 10$ antennas and the user with $N_r = 2$ antennas. The transmission power levels are varied across $P \in \{20, 30, 40, 50, 60\}$ dBm, while the noise variance is fixed at $\sigma^2 = -80$ dBm. The RIS is modeled as a UPA with a total of $N = 64$ elements, constructed from vertical and horizontal dimensions where $N_v = N_h = 8$. The RIS gain is set to $G_{\rm RIS} = 5$ dB. We adopt a practical model that captures phase-dependent amplitude distortion with the following parameters: minimum amplitude $\beta_{\min} = 0.2$, shaping parameter $\alpha = 1.6$, and minimum-gain phase $\psi_0 = 0.43\pi$. For the channel model, the environment includes $L = 2$ NLOS paths, and the Rician factors for the BS-to-RIS and RIS-to-user links are both set to $K_t = K_r = 10$. The path loss exponent and distance for the BS-RIS link are $\xi_t = 2$ and $d_t = 10$ meters, while for the RIS-user link, the values are $\xi_r = 2.8$ and $d_r = 30$ meters, respectively.

The LSTM-based network is trained using $10^6$ samples for the training set, with $10\%$ reserved for validation and $10^3$ samples for testing. The batch size is set to $2 \times 10^3$, and the learning rate is initialized at $10^{-2}$. The Adam optimizer is adopted to accelerate convergence and improve generalization. An early stopping criterion of 2 epochs is applied to prevent overfitting. To reduce the signaling overhead, a structured RIS codebook is constructed by rearranging DFT beam-steering vectors across quantized azimuth and elevation directions.

To evaluate the performance of the proposed LSTM model, we compare it against four approaches:
\begin{enumerate}
    \item \textbf{ES (upper bound)}: evaluates all possible codewords in the DFT codebook.
    \item \textbf{AO}: iterative baseline~\cite{Ruijin2023}, up to 30 iterations per realization.
    \item \textbf{CNN}: two 2D convolutional layers (16 filters, size $2 \times 2$, Leaky ReLU), a flattening layer, two fully connected hidden layers (256 and 128 neurons), and a Sigmoid output layer with 64 neurons.
    \item \textbf{Random Selection (lower bound)}: selects a codeword uniformly at random.
\end{enumerate}

\begin{figure}[t]
\centering
{\includegraphics[width=0.35\textwidth]{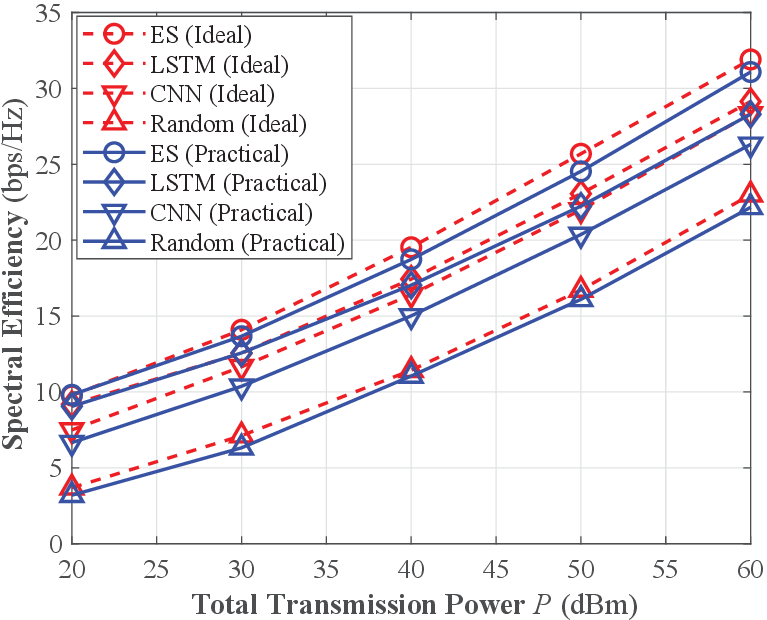}}
\caption{Spectral efficiency versus transmission power, comparing LSTM, CNN, ES, AO, and random selection ($K_t = K_r = 10$, $N_{t} = 10$, $N_{r} = 2$, $N = 64$, and $L = 2$).}
\label{LSTM_CNN_ES}
\vspace{-0.1in}
\end{figure}

Fig.~\ref{LSTM_CNN_ES} shows the spectral efficiency under both ideal and practical RIS reflection models. The proposed LSTM consistently outperforms CNN across all transmit power levels. Under the practical model, LSTM achieves between $90.6\%$ and $92.2\%$ of ES, outperforming CNN by up to $35.8\%$. The AO baseline achieves $94.5\%$ of ES but requires more than 25 ms per channel realization, making it impractical for real-time use. In contrast, LSTM maintains near-optimal performance at only $0.59$ ms inference time.

\begin{figure}[t]
\centering
{\includegraphics[width=0.35\textwidth]{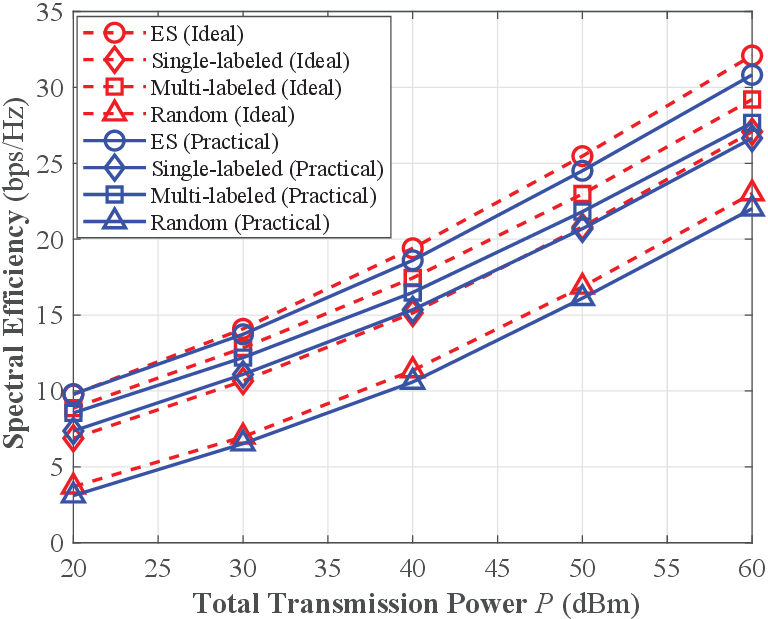}}
\caption{Spectral efficiency versus transmission power, comparing single-labeled and multi-labeled LSTM under ideal and practical RIS models.}
\label{Labels}
\end{figure}

Fig.~\ref{Labels} compares the proposed multi-labeled strategy with single-labeled LSTM training. Under the ideal RIS model, multi-label LSTM gains $7.8\%$ to $29\%$ improvement across power levels. Under the practical model, it improves $3.8\%$ to $16.1\%$, achieving up to $89.7\%$ of ES. This confirms the benefit of considering multiple near-optimal codewords to mitigate the non-convexity of RIS optimization.

\begin{table}[t]
   \caption{Inference time and spectral efficiency normalized to ES under practical RIS constraints.}
   \label{mmWave_comparison}
   \scriptsize
   \centering
   \begin{tabular}{|l|c|c|c|c|c|}
   \hline
   \textbf{Schemes} & ES & AO~\cite{Ruijin2023} & LSTM & CNN & Random\\ 
   \hline
   Computation time (ms)  &26.75 &12.10 &0.59  &4.45 &0.048\\
   \hline
   Spectral efficiency (\%) &100	&94.5 &92.2	&67.9 &32.6\\
   \hline
   \end{tabular}
   \vspace{-0.05in}
\end{table}

Table~\ref{mmWave_comparison} shows that the proposed LSTM achieves $92.2\%$ of ES with only $2.2\%$ of its computation time. Compared with AO, LSTM reduces latency by over 97\% while maintaining a performance gap of less than 3\%.

\begin{table}[t]
\centering
\caption{Computational complexity comparison of different schemes.}
\label{tab:complexity}
\scriptsize
\begin{tabular}{|l|c|c|}
\hline
\textbf{Scheme} & \textbf{Complexity Order} & \textbf{Remarks} \\ \hline
ES & $\mathcal{O}(Q \cdot N_t N_r)$ & Optimal but infeasible in real time \\ \hline
AO~\cite{Ruijin2023} & $\mathcal{O}(I \cdot Q \cdot N_t N_r)$ & $I$: iterations, high latency \\ \hline
CNN & $\mathcal{O}(K^2)$ & 2D convolutions, slower inference \\ \hline
LSTM & $\mathcal{O}(K \cdot H)$ & One forward pass, sustainable \\ \hline
\end{tabular}
\vspace{-0.2in}
\end{table}

Table~\ref{tab:complexity} shows that the proposed LSTM reduces the complexity order from $\mathcal{O}(Q \cdot N_t N_r)$ for ES to $\mathcal{O}(K \cdot H)$, requiring only a single forward pass. This advantage becomes more pronounced for large RIS arrays, where $Q = N_h \times N_v$ grows rapidly.

\textbf{Robustness Analysis:} We further evaluate performance under perturbed propagation parameters, varying path loss exponents $\xi_{t}, \xi_{r}$ and Rician factors $K_t, K_r$ by $\pm 20\%$ from training settings. The LSTM maintains above $88\%$ of ES performance, confirming its resilience to distribution mismatch.

\textbf{Sustainability Discussion:} Following the conversion ratio $1$ GFLOP $\approx 0.1$ joules for advanced baseband processors~\cite{Zhang2024MCOM}, the proposed approach reduces inference energy consumption by nearly two orders of magnitude compared with ES or AO. This aligns with the objectives of sustainable AI-driven wireless systems.

\section{Conclusion}
We proposed a sustainable LSTM-based precoding framework for RIS-assisted mmWave MIMO systems that leverages uplink pilot sequences to bypass explicit CSI estimation. By incorporating the practical phase-dependent amplitude model and a multi-label training strategy, the scheme achieves over $90\%$ of exhaustive search performance with only $2.2\%$ of its computation time. 
The model sustains above $88\%$ performance under distribution mismatch and scales effectively to larger RIS arrays with marginal latency increases, while significantly reducing energy consumption. These results demonstrate its potential for real-time and energy-efficient 6G deployments. Future work will extend the framework to multi-user and high-mobility scenarios.

\ifCLASSOPTIONcaptionsoff
  \newpage
\fi

\end{document}